\documentclass{svjour3}
\usepackage{graphicx}
\usepackage{epsfig}
\usepackage{amssymb,amsmath,scalefnt}

\begin{document}

\title{Generalized Kraus operators for the one-qubit
depolarizing quantum channel}

\author{M. Arsenijevi\' c        \and    J. Jekni\' c-Dugi\' c   \and  M. Dugi\' c}

\institute{M. Arsenijevi\' c \at
              University of Kragujevac, Faculty of Science
, Kragujevac 34000, Serbia \\
             \email{momirarsenijevic@kg.ac.rs}
             \and
           J. Jekni\' c-Dugi\' c \at
           Department of Physics, Faculty of Science and Mathematics, University of Ni\v s, Ni\v s 18000, Serbia \\
            \email{jjeknic@pmf.ni.ac.rs}
              \and
              M. Dugi\' c \at
              University of Kragujevac, Faculty of Science, Kragujevac 34000, Serbia \\
           \email{dugic@kg.ac.rs}
}

\date{Received: date / Accepted: date}

\maketitle

\begin{abstract}
Microscopic Hamiltonian models of the composite system ``open
system + environment'' typically do not provide the operator-sum
Kraus form of the open system's dynamical map. With the use of a
recently developed method \cite{AndCresHall}, we derive the Kraus
operators starting from the microscopic Hamiltonian model, i.e.
from the proper master equation, of the one-qubit depolarizing
channel. Those Kraus operators  generalize the standard
counterparts, which are widely used in the literature. Comparison
of the standard and the here obtained Kraus operators is performed
via investigating dynamical change of the Bloch sphere volume,
entropy production and the open system's state trace distance. We compare the generalized with the standard Kraus operators for both single qubit as well as regarding the occurrence of the entanglement sudden death for a pair of initially correlated qubits. We find that the generalized Kraus operators describe the less deteriorating quantum channel than the standard ones.
 \keywords{Open quantum systems \and
Kraus operator-sum decomposition \and Qubit operations}
 \PACS{03.65.Yz \and 03.65.Ta \and 03.65.-w \and 03.67.Mn \and 03.65.Yz \and 65.40G-}
\end{abstract}

\section{Introduction}\label{Introduction}

The differential master equations and the integral operator-sum
Kraus form of the completely positive dynamical maps are the basic
however mutually non-equivalent methods in modern open quantum
systems theory \cite{KrausSEO,BreuPetr,RivasHuelga,N&Ch}. On the
one hand, for the completely positive (CP) dynamical maps,
solution of a master equation can always be expressed as a quantum
operation \cite{KrausSEO} in the operator-sum representation
\cite{BreuPetr,RivasHuelga,N&Ch}. On the other hand, an
operator-sum representation cannot necessarily be presented in a
master-equation form \cite{BreuPetr,RivasHuelga,N&Ch,PRSA}. Hence the
relative advantage of the operator-sum {\it formalism}.

Nevertheless, in the case of the Markovian (and particularly
time-ho\-mo\-gen\-eous semigroup) dynamics
\cite{BreuPetr,RivasHuelga,N&Ch}, when the methods are
interchangeable, there is a {\it physical} advantage of the
master-equations formalism that is worth emphasizing. Actually,
the master equations formalism fully respects the microscopic
details in the Hamiltonian form of the composite system ``open
system + environment''. These details regard the realistic
physical situations and, in principle, open the door for the
better control of the open system's dynamics. However, such
microscopic-model details are typically absent from the
operator-sum representation of the open-system's dynamics. That
is, the Kraus operators are often constructed without the clear
and unambiguous microscopic-model criteria \cite{N&Ch}. Hence the
natural, as yet virtually intact, question arises: whether or not
the widely used Kraus operators \cite{N&Ch} reliably represent the
microscopically modelled quantum noisy channels?

In this paper we answer this question for the quantum depolarizing
channel
\cite{N&Ch,RomLoFranco,BCSQIQC,QCD,ByChManiscalco,Marinescu,KlimSoto,GeomEntDurstPhD}
while the analogous results for the (generalized) amplitude and
phase damping processes can be found in \cite{KJS}.  We make use
of a recently devised method for deriving the Kraus operators from
a local-in-time master equation for the finite-dimensional open
systems \cite{AndCresHall}. Formally, this procedure is equivalent
to solving the master equation, whether Markovian, non-Markovian,
of Lindblad form or not. The method can also be used in the
opposite direction.

We first derive an interaction-picture master equation and hence
derive the \textit{novel} Kraus operators that generalize the
standard ones. The new Kraus operators can be termed
``generalized'' as the proper choice of the model parameters
reduce the new to the standard Kraus operators for the depolarizing  (DP) channel.
Dissent of the novel from the standard Kraus operators is
investigated via temporal behavior of the Bloch sphere volume, the
state trace-distance and entropy production. We find that, in
short time intervals, the generalized DP channel is less
deteriorating than the standard DP channel, while in the
asymptotic limit the channels are indistinguishable. This finding is confirmed by investigating entanglement dynamics for a pair of qubits initially in a maximally entangled state.

Structure of this paper is as follows. In Section \ref{The method}
we overview the basics of the procedure devised in
Ref.\cite{AndCresHall}. Based on a master equation for the
depolarizing channel that is derived in Appendix, in Section
\ref{KrausDP} we derive the  generalized Kraus operators for the
depolarizing channel. In Section \ref{Analysis} we perform
comparison of the standard and the here obtained generalized
depolarizing processes. Section \ref{Discussion} is discussion and
we conclude in Section \ref{Conclusion}.

\section{Overview of the method}\label{The method}
In Ref. \cite{AndCresHall}, the authors  developed a general
procedure for deriving a Kraus decomposition from a known master
equation and vice versa, regarding the  finite-dimensional quantum
systems. The only assumption is that the master equation is local
in time.

The so-called Nakajima-Zwanzig projection method
\cite{BreuPetr,RivasHuelga} gives the following master equation
for the system's density operator $\hat{\rho}_S(t)$, ($\hbar=1$):
\begin{equation}\label{NZmaster}
\frac{d\hat{\rho}_S(t)}{dt}=-i[\hat{H},\hat{\rho}_S(t)]+\int_{0}^{t}
\mathcal{K}_{t,s}[\hat{\rho}_S(s)]ds,
\end{equation}
where $\hat{H}$ represents the system's self-Hamiltonian (that
includes the so-called Lamb-shift term) and $\mathcal{K}_{t,s}$ is
the memory kernel which accounts for the non-unitary effects  due
to  the environment.

Certain processes can be written  in a local-in-time form
\cite{BreuPetr,RivasHuelga,ChatShi,ShiTakash}:
\begin{equation}\label{Localmaster}
\dot{\hat{\rho}}_{S}(t)=\Lambda_{t}(\hat{\rho}_{S}(t)),
\end{equation}
where $\Lambda_{t}$ is a linear map which preserves hermiticity,
positivity and  unit trace of $\hat{\rho}_{S}(t)$ and has the
property:
\begin{equation}\label{Traceproperty}
tr\Lambda_{t}(\hat{\rho}_{S}(t))=0.
\end{equation}

Alternatively, dynamics can be presented in a non-differential,
``integral'' form \cite{BreuPetr,RivasHuelga,AndCresHall}:
\begin{equation}\label{Integralform}
\hat{\rho}_{S}(t)=\phi_{t}(\hat{\rho}_{S}(0)),
\end{equation}
where $\phi_{t}$ is a completely positive and trace preserving,
linear map.

It can be shown \cite{AndCresHall}, that the linear maps
$\Lambda_{t}$ and $\phi_{t}$ are connected via the matrix
differential equation:
\begin{equation}\label{IntDiffconnection}
\dot{F}=LF,
\end{equation}
where the $F$ matrix represents the $\phi_t$ map while the matrix
elements of $L$ are given by:
\begin{equation}\label{MatrixelemL}
L_{kl}=tr[\mathcal{G}_k\Lambda(\mathcal{G}_l)].
\end{equation}
In eq.\eqref{MatrixelemL}, $\{\mathcal{G}_k\}$ is  an orthonormal
basis  of the Hermitian operators acting on the system's Hilbert
state space. For the time independent $\Lambda_t$, i.e. $L$,
eq.\eqref{IntDiffconnection} has the unique solution:
\begin{equation}\label{Integralformsolution}
F=e^{Lt}.
\end{equation}

Complete positivity of the map $\phi_t$ (and hence of the matrix
$F$) is equivalent to the positivity of the, so called, Choi
matrix, $S$ \cite{AndCresHall,Choi} whose elements are defined
as \cite{AndCresHall}:
\begin{equation}\label{Choimatrix}
S_{nm}=\sum_{s,r}F_{sr}tr[\mathcal{G}_r\mathcal{G}_n^{\dag}\mathcal{G}_s\mathcal{G}_m],
\end{equation}

\noindent where $F_{sr}$s are entries of the $F$ matrix,
eq.\eqref{IntDiffconnection}.

With the use of eq. \eqref{Choimatrix}, eq.\eqref{Integralform}
takes the form:
\begin{equation}\label{AlternDecomp}
\phi(\hat{\rho}_S(0))=\sum_{nm}S_{nm}\mathcal{G}_n\hat{\rho}_S(0)\mathcal{G}_m^{\dag},
\end{equation}
which, after diagonalization of the  $S$ matrix:
\begin{equation}\label{ChoiDiag}
S=UDU^{\dag},
   \end{equation}
gives rise to a Kraus decomposition. The eigenvalues $d_i$ and the
eigenvectors of the $S$ matrix constitute the diagonal matrix $D$
and the unitary matrix $U=(u_{ij})$, respectively; columns of the
unitary matrix $U$  are  the normalized eigenvectors of the $S$
matrix. Then the Kraus operators:
\begin{equation}\label{KrausoperViaDandU}
E_i=\sum_j\sqrt{d_i}u_{ji}\mathcal{G}_j
   \end{equation}
yield the Kraus decomposition of the  dynamical map $\phi_t$:
\begin{equation}\label{KrausDecompphiB}
\phi_t(\hat{\rho}_S(0))=\sum_{k}\hat{E}_k(t)\hat{\rho}_S(0)\hat{E}_k^{\dag}(t).
\end{equation}

Hence the chain of the construction is established: from a master
equation to calculate $L$, then via relation
\eqref{Integralformsolution} to obtain the matrix $F$ and, due to
eq.\eqref{Choimatrix} and eq.\eqref{ChoiDiag}, to calculate the
Kraus operators, eq.\eqref{KrausoperViaDandU}.

\section{The depolarizing channel}\label{DPchannel}

While microscopic derivation for the phase and amplitude damping
master equations can be found in the  literature
\cite{RivasHuelga}, to the best of our knowledge, this is not the
case for the master equation concerning the depolarizing process,
eq.\eqref{RomLoFrancoDPmaster}. One possible form (due to the
unitary freedom) of the standard depolarizing Kraus operators is
given by\cite{N&Ch,RomLoFranco,BCSQIQC,Marinescu}:
\begin{align}\label{RomLoFrancoDPKraus}
\hat{E}_{0}&=\sqrt{1-\frac{3p(t)}{4}} \hat{I},&
\hat{E}_{i}&=\frac{\sqrt{p(t)}}{2}\hat{\sigma}_{i}, (i=x,y,z)
\end{align}

\noindent with the associated, {\it phenomenological} master
equation in the interaction picture:

\begin{align}\label{RomLoFrancoDPmaster}
\frac{d\hat{\rho}_S(t)}{dt}&=\frac{\gamma}{8}\sum_i(
   2\hat{\sigma}_i\hat{\rho}_S(t)\hat{\sigma}_i-\hat{\sigma}_i^2\hat{\rho}_S(t)-
   \hat{\rho}_S(t)\hat{\sigma}_i^2)\,,
\end{align}
where  the Pauli sigma-operators  $\hat{\sigma_i}$ ($i=x,y,z$)
pertain to the qubit's degrees of freedom and
 $\gamma$ is so-called damping
constant, while $p(t)=1-e^{-\gamma t}$.

In this section, for a microscopic (Hamiltonian)  model we derive
the Kraus operators starting from a Markovian (Lindblad-form)
master equation, which generalizes eq.\eqref{RomLoFrancoDPmaster}
and therefore leads to a generalization of the standard Kraus
operators, eq.\eqref{RomLoFrancoDPKraus}. Thereby we implicitly
test derivability of eq.\eqref{RomLoFrancoDPKraus} from
eq.\eqref{RomLoFrancoDPmaster} that here will not be made
explicit. Analogous derivation regarding the GAD and PD channels
can be found in \cite{KJS}.

The microscopic model for  the depolarizing  process we are
interested in is given by the following Hamiltonian
\cite{RomLoFranco}:
\begin{equation}\label{A1}
    \hat{H}=\hat{H}_S\otimes\hat{I}_E+\hat{I}_S\otimes\hat{H}_E+\hat{A}_x\otimes\hat{B}_x
    +\hat{A}_y\otimes\hat{B}_y+\hat{A}_z\otimes\hat{B}_z.
\end{equation}

The terms in eq.\eqref{A1} [in the physical units of $\hbar = 1$]:
the self-Hamiltonians $\hat{H}_S=\frac{\omega_0}{2}\hat{\sigma}_z$
and $\hat{H}_E=\sum_i\int^{\omega_{max}}_0 d\omega
\hat{a}^{\dag}_{\omega i}\hat{a}_{\omega i}$ for the $S$ system
and the environment (thermal bath) $E$, respectively, while the
interaction terms
$\hat{A}_i\otimes\hat{B}_i=\hat{\sigma}_i\otimes\int^{\omega_{max}}_0
d\omega h(\omega) (\hat{a}^{\dag}_{\omega i}+\hat{a}_{\omega i})$,
$i=x,y,z$, with the coupling functions $h(\omega)$ accounting for
the interaction strength per frequency. From eq.\eqref{A1} we can
learn that the environment $E$ effectively acts as a set of three
independent bosonic environments, $E_i, i=x,y,z$, with the
standard, Bose-operators,  commutation relations $[\hat{a}_{\omega
i},\hat{a}_{\omega' j}^{\dag}]=\delta(\omega-\omega')\delta_{ij}$,
$i,j=x,y,z$.

Our approach is minimalist in the sense that we do not introduce
assumptions that might force us to leave the standard physical
models. Therefore we are concerned with the time-homogeneous $L$,
cf. eq.\eqref{MatrixelemL}, as well as with the Ohmic spectral
density.

\subsection{Kraus operators for the generalized depolarizing channel}\label{KrausDP}

In Appendix we derive the following, Markovian (Lindblad-form),
master equation for the microscopic model eq.\eqref{A1} of the
depolarizing channel:

\begin{equation}\label{GeneralizedMJ}
       \begin{split}
\displaystyle\frac{d\hat{\rho}_S(t)}{dt}&=
-i\Delta[\hat{\sigma}_z,\hat{\rho}_S(t)]+ \gamma_{zz}(0,
T)(\hat{\sigma}_{z}\hat{\rho}_S(t)\hat{\sigma}_{z}-\hat{\rho}_S(t))+\\&
+\frac{\gamma_{xx}(\omega_0, T)+\gamma_{yy}(\omega_0,
T)}{2}(\hat{\sigma}_{x}\hat{\rho}_S(t)
\hat{\sigma}_{x}-\hat{\rho}_S(t))+\\& +\frac{\gamma_{xx}(\omega_0,
T)+\gamma_{yy}(\omega_0, T)}{2}
(\hat{\sigma}_{y}\hat{\rho}_S(t)\hat{\sigma}_{y}-\hat{\rho}_S(t))\,.
\end{split}
\end{equation}

\noindent Definitions of the constants appearing in
eq.\eqref{GeneralizedMJ} can be found in Appendix where we
strongly emphasize  that the microscopic model eq.\eqref{A1}
allows for the Markovian dynamics eq.\eqref{GeneralizedMJ}
\textit{only if} the spectral density is of the Ohmic type and
$\omega_0/\omega_{c} \ll 1$, while the high temperature limit
gives rise to the constraint $k_B T/\hbar\omega_0 \gg 1$ and
$\omega_{c}$ is the cutoff frequency.

Eq.\eqref{GeneralizedMJ}  straightforwardly reduces to the
standard one, eq.\eqref{RomLoFrancoDPmaster}, for $\Delta=0$ (no
Lamb shift term) and $2
\gamma_{zz}(0,T)=\gamma_{xx}(\omega_0,T)+\gamma_{yy}(\omega_0,T)$.
From eq.\eqref{A1} it follows the same interaction-strength per
frequency $h(\omega)$, which givers rise the same interaction
strength, $\alpha$ for all environments--see Appendix.

To ease the calculation we  introduce: $\mathbf{x}=\Delta$,
$\mathbf{y}=\gamma_{zz}$ and
$\mathbf{z}=\frac{\gamma_{xx}+\gamma_{yy}}{2}$ with which the
equation (16) reads:
\begin{equation}\label{DPmastereqxyz}
       \begin{array}{r c l}
\displaystyle\frac{d\hat{\rho}_S(t)}{dt}&=&
-i\mathbf{x}[\hat{\sigma}_z,\hat{\rho}_S(t)]+
\mathbf{y}(\hat{\sigma}_{z}\hat{\rho}_S(t)\hat{\sigma}_{z}-\hat{\rho}_S(t))\\&
+&\mathbf{z}(\hat{\sigma}_{x}\hat{\rho}_S(t)\hat{\sigma}_{x}-\hat{\rho}_S(t))+
\mathbf{z}(\hat{\sigma}_{y}\hat{\rho}_S(t)\hat{\sigma}_{y}-\hat{\rho}_S(t))\,,
\end{array}
\end{equation}

Then, the use of eq.\eqref{MatrixelemL} gives rise to:

\begin{equation}\label{DPmatL}
L=\left(
\begin{array}{cccc}
 0 & 0 & 0 & 0 \\
 0 & -2 (\mathbf{y}+\mathbf{z}) & -2 \mathbf{x} & 0 \\
 0 & 2 \mathbf{x} & -2 (\mathbf{y}+\mathbf{z}) & 0 \\
 0 & 0 & 0 & -4 \mathbf{z}
\end{array}
\right)\,.
\end{equation}

Multiplication of the matrix $L$ by
$\frac{1}{2(\mathbf{y}+\mathbf{z})}$ and introduction of the new
parameters, $\theta =\frac{\mathbf{x}}{\mathbf{y}+\mathbf{z}}$,
$\Omega= -\frac{2 \mathbf{z}}{\mathbf{y}+\mathbf{z}}$, $\tau
=2(\mathbf{y}+\mathbf{z})t$, give

\begin{equation}\label{DPmatLthetaOmegatau}
L'=  \frac{L}{2(\mathbf{y}+\mathbf{z})}=\left(
\begin{array}{cccc}
 0 & 0 & 0 & 0 \\
 0 & -1 & -\theta  & 0 \\
 0 & \theta  & -1 & 0 \\
 0 & 0 & 0 & \Omega
\end{array}
\right)\,
\end{equation}
while $\Omega\in(-2,0)$, $\tau\in(-\infty,\infty)$.

Now eq.\eqref{Integralformsolution} can be written as:

\begin{equation}\label{DPmatF}
 F=e^{L'\tau},
\end{equation}

\noindent that makes easy the calculation of the exponential
matrix $F$:

\begin{equation}\label{DPmatF1}
F=\left(
\begin{array}{cccc}
 1 & 0 & 0 & 0 \\
 0 & e^{-\tau } \cos (\theta  \tau ) & -e^{-\tau } \sin (\theta  \tau ) & 0 \\
 0 & e^{-\tau } \sin (\theta  \tau ) & e^{-\tau } \cos (\theta  \tau ) & 0 \\
 0 & 0 & 0 & e^{\tau  \Omega }
\end{array}
\right)\,.
\end{equation}

\noindent as well as of the  Choi matrix, cf.
eq.\eqref{Choimatrix}:
\begin{equation}\label{DPmatS}
  S=\left(\scalefont{0.8}{
\begin{array}{cccc}
 e^{-\tau } \cos (\theta  \tau )+\frac{e^{\tau  \Omega }}{2}+\frac{1}{2} &
 0 & 0 & i e^{-\tau } \sin (\theta  \tau ) \\
 0 & \frac{1}{2}-\frac{e^{\tau  \Omega }}{2} & 0 & 0 \\
 0 & 0 & \frac{1}{2}-\frac{e^{\tau  \Omega }}{2} & 0 \\
 -i e^{-\tau } \sin (\theta  \tau ) & 0 & 0 & -e^{-\tau } \cos (\theta  \tau )+
 \frac{e^{\tau  \Omega }}{2}+\frac{1}{2}
\end{array}}
\right)\,.
\end{equation}

Diagonalization of the $S$ matrix, eq.\eqref{DPmatS}, gives the
eigenvalues:
\begin{subequations}\label{DPEigenvalues}
\begin{align}
\frac{1}{2} \left(1-e^{\tau  \Omega }\right), \\
\frac{1}{2} \left(1-e^{\tau  \Omega }\right), \\[0.5ex]
\frac{1}{2} e^{-\tau } \left(-2+e^{\tau }+e^{\tau +\tau  \Omega
}\right),\\ \frac{1}{2} e^{-\tau } \left(2+e^{\tau }+e^{\tau +\tau
\Omega }\right)\,,
\end{align}
\end{subequations}

\noindent and  the  respective non-normalized eigenvectors: 
\begin{subequations}\label{DPEigenvectors}
\center{
\begin{align}
 \{0,0,1,0\},\\  \{0, 1, 0, 0\},\\ \{i (\cot[\theta  \tau ]-\csc[\theta  \tau ]),0,0,1\},
 \\\{i (\cot[\theta  \tau ]+\csc[\theta  \tau ]),0,0,1\}\}\,.
\end{align}}
\end{subequations}

With the use of eq.\eqref{KrausoperViaDandU}, from
eqs.\eqref{DPEigenvalues}-\eqref{DPEigenvectors} follow the
desired Kraus matrices:
\begin{equation}\label{DP1}
\mathbb{E}_1=\left(
\begin{array}{cc}
 0 & -\frac{1}{2} i \sqrt{1-e^{\tau  \Omega }} \\
 \frac{1}{2} i \sqrt{1-e^{\tau  \Omega }} & 0
\end{array}
\right)
\end{equation}

\begin{equation}\label{DP2}
\mathbb{E}_2=\left(
\begin{array}{cc}
 0 & \frac{1}{2} \sqrt{1-e^{\tau  \Omega }} \\
 \frac{1}{2} \sqrt{1-e^{\tau  \Omega }} & 0
\end{array}
\right)
\end{equation}

\begin{equation}\label{DP3}
\mathbb{E}_3=\left(
\begin{array}{cc}
 \frac{1}{2} \left(1-i \tan \left(\frac{\theta  \tau }{2}\right)\right)
 \sqrt{\frac{e^{\tau  \Omega }-2 e^{-\tau }+1}{\tan ^2\left(\frac{\theta  \tau }{2}\right)+1}} & 0 \\
 0 & \frac{1}{2} \left(-1-i \tan \left(\frac{\theta  \tau }{2}\right)\right)
 \sqrt{\frac{e^{\tau  \Omega }-2 e^{-\tau }+1}{\tan ^2\left(\frac{\theta  \tau }{2}\right)+1}}
\end{array}
\right)
\end{equation}

\begin{equation}\label{DP4}
\mathbb{E}_4=\left(
\begin{array}{cc}
 \frac{1}{2} \left(1+i \cot \left(\frac{\theta  \tau }{2}\right)\right)
 \sqrt{\frac{e^{\tau  \Omega }+2 e^{-\tau }+1}{\cot ^2\left(\frac{\theta  \tau }{2}\right)+1}} & 0 \\
 0 & \frac{1}{2} i \left(\cot \left(\frac{\theta  \tau }{2}\right)+i\right)
 \sqrt{\frac{e^{\tau  \Omega }+2 e^{-\tau }+1}{\cot ^2\left(\frac{\theta  \tau }{2}\right)+1}}
\end{array}
\right)
\end{equation}

It is easy to show that the  Kraus matrices
\eqref{DP1}-\eqref{DP4} satisfy the completeness relation $
\sum_{k}\hat{\mathbb{E}}_k(t)^{\dag}\hat{\mathbb{E}}_k(t)=\hat{I}$
and also reduce to the standard ones in
eq.\eqref{RomLoFrancoDPKraus} with the following choice of the
parameters: $\theta=0$ (equivalently $\mathbf{x}=0$, neglecting
the Lamb shift) and $\Omega=-1$ (equivalently
$\mathbf{y}=\mathbf{z}$).  That is, for this choice of the
parameters values, from eqs. \eqref{DP1}-\eqref{DP4} easily follow
approximations of the Kraus operators eqs.\eqref{DP1}-\eqref{DP4},
respectively:

\begin{equation}\label{DP1Stand}
\mathbb{E}_1=\left(
\begin{array}{cc}
 0 & -\frac{1}{2} i \sqrt{1-e^{-\tau }} \\
 \frac{1}{2} i \sqrt{1-e^{-\tau }} & 0
\end{array}
\right)\,,
\end{equation}

\begin{equation}\label{DP2Stand}
\mathbb{E}_2=\left(
\begin{array}{cc}
 0 & \frac{1}{2} \sqrt{1-e^{-\tau }} \\
 \frac{1}{2} \sqrt{1-e^{-\tau }} & 0
\end{array}
\right)\,,
\end{equation}

\begin{equation}\label{DP3Stand}
\mathbb{E}_3=\left(
\begin{array}{cc}
 \frac{1}{2} \sqrt{1-e^{-\tau }} & 0 \\
 0 & -\frac{1}{2} \sqrt{1-e^{-\tau }}
\end{array}
\right)\,,
\end{equation}

\begin{equation}\label{DP4Stand}
\mathbb{E}_4=\left(
\begin{array}{cc}
 -\frac{1}{2} i e^{-\frac{\tau }{2}} \sqrt{3+e^{\tau }} & 0 \\
 0 & -\frac{1}{2} i e^{-\frac{\tau }{2}} \sqrt{3+e^{\tau }}
\end{array}
\right)\,.
\end{equation}

The matrices \eqref{DP1Stand}-\eqref{DP3Stand}  are the
$\hat{\sigma}_z=|0\rangle\langle0|-|1\rangle\langle1|$
representation of the standard Kraus operators in
eq.\eqref{RomLoFrancoDPKraus} while the matrix $\mathbb{E}_4'=2i
\mathbb{E}_4$ exhibits a unitary freedom for the Kraus operator
$\mathbb{E}_4$ in eq.\eqref{RomLoFrancoDPKraus}.

\subsection{Comparison of the generalized and the standard depolarizing channels}\label{Analysis}

The Kraus matrices eqs.\eqref{DP1}-\eqref{DP4} represent a
generalization of the standard DP channel
eq.\eqref{RomLoFrancoDPKraus} (i.e.
\eqref{DP1Stand}-\eqref{DP4Stand}). In this section we compare
these two kinds of the noisy channels by comparing their dynamical
effects regarding the Bloch-sphere volume, the von Neumann entropy
and  the trace distance. In order to give precise meaning of the
task, we first provide definitions of these standard
quantum-information quantities.

The Bloch sphere is a geometrical representation of the
single-qubit state space where every pure state  can be written as
\cite{N&Ch}

\begin{equation}\label{DensOperBloch}
   \hat{\rho}=\frac{1}{2}\left(\hat{I} +\vec{n} \cdot \hat{\vec{\sigma}}
   \right)
\end{equation}

\noindent where the vector $\vec{n}$ of the unit length, $\vert
\vec{n}\vert = 1$, corresponds to the points on the Bloch sphere
determined by the spherical coordinates $\vec{n}=(\sin v \cos u,
\; \sin v  \sin u, \; \cos v)$ and uniquely determine a pure
quantum state of a single qubit; $v\in[0,\pi]$, $u\in[0,2\pi]$.
The points inside the sphere correspond one-to-one to the mixed
states for every $\vert\vec{n}\vert < 1 $. The noisy channel
effects can be geometrically presented as the Bloch-sphere
deformation \cite{N&Ch}. Denoting the volume of the [deformed]
Bloch sphere $V(t)$ in an instant of time $t$, the relative speed
of the volume-change is defined as:

\begin{equation}\label{Volumevariation}
   \kappa(t)=\frac{1}{V_0}
\frac{dV(t)}{dt}
\end{equation}

\noindent where $V_0$ is the volume at $t=0$.

The von Neumann entropy of a quantum state $\hat\rho$ in an
instant $t$ of time

\begin{equation}\label{vonNeumannEntropy}
    S(t) = - \mathrm{tr}(\hat{\rho}(t) \ln\hat{\rho}(t)),
\end{equation}

\noindent increases for all kinds of the noisy channels
\cite{N&Ch}. Increase of the von Neumann entropy is an alternative
(non-geometrical) description of the Bloch-sphere deformation that
brings the information-theoretic description of the regarded
processes \cite{N&Ch} that cannot be uniquely deduced from
eq.\eqref{Volumevariation}.

The so-called trace distance

\begin{equation}\label{TraceDistance}
 T(\hat{\rho},\hat{\sigma}) := \frac{1}{2}||\hat{\rho} - \hat{\sigma}||_{1} =
 \frac{1}{2} \mathrm{tr}\left[\sqrt{(\hat{\rho}-\hat{\sigma})^\dagger (\hat{\rho}-\hat{\sigma})}
 \right]
\end{equation}

\noindent quantifies how much the quantum states $\hat\rho$ and
$\hat\sigma$ differ from each other \cite{N&Ch}. Hence comparison
of the regarded channels is possible with the use of
eq.\eqref{TraceDistance} by setting $\hat\rho(t)$ and
$\hat\sigma(t)$ as the states produced by the generalized and by
the standard DP channel, respectively.

For the generalized DP channel, i.e. from eqs.
(\ref{DP1}-\ref{DP4}) follows:
\begin{subequations} \label{DPEffects}
\begin{align}
\phi_{\tau}(\hat{I})&=\hat{I}\,, \label{DPEffectsA} \\
\phi_{\tau}(\hat{\sigma}_x)&=e^{-\tau}(\hat{\sigma}_x\cos\theta\tau+\hat{\sigma}_y\sin\theta\tau)\,, \label{DPEffectsB}\\
\phi_{\tau}(\hat{\sigma}_y)&=e^{-\tau}(\hat{\sigma}_y\cos\theta\tau-\hat{\sigma}_x\sin\theta\tau)\,, \label{DPEffectsC}\\
\phi_{\tau}(\hat{\sigma}_z)&=e^{\tau\Omega}\hat{\sigma}_z\,,
\label{DPEffectsD}
\end{align}
\end{subequations}
\noindent that is
\begin{equation}\label{DPDensOperBlochthetaOmegatau}
\begin{split}
  \phi_{\tau}(\hat{\rho})&=\frac{1}{2}[\hat{I}+e^{-\tau}\sin v\cos(u+\theta\tau)\hat{\sigma}_x+\\&
  +e^{-\tau}\sin v\sin(u+\theta\tau)\hat{\sigma}_y+e^{\tau\Omega}\cos
  v\hat{\sigma}_z]\,,
  \end{split}
\end{equation}
\noindent and equivalently:
\begin{equation}\label{DPDensOperBlochxyz}
\begin{split}
 \phi_{t}(\hat{\rho})&=\frac{1}{2}[\hat{I}+e^{-2(\mathbf{y}+\mathbf{z})t}\sin v
  \cos(u+2\mathbf{x}t)\hat{\sigma}_x+\\&
  +e^{-\tau}\sin v\sin(u+2\mathbf{x}t)\hat{\sigma}_y+e^{-4\mathbf{z}t}\cos
  v\hat{\sigma}_z]\,.
  \end{split}
\end{equation}
\noindent Expressions \eqref{DPDensOperBlochthetaOmegatau} and
\eqref{DPDensOperBlochxyz} are solutions of the master equation
eq.\eqref{GeneralizedMJ}. It is well known that for Pauli channel
Pauli matrices are eigenvectors of the channel, but, as can be
seen from eqs.\eqref{DPEffects}, it is not the case for
generalized depolarizing (GDP) channel.

From eq.\eqref{DPDensOperBlochxyz} we analytically obtain the time
dependence

\begin{equation}\label{DPBlochvolumechange}
   V(\tau)=\frac{4\pi}{3}e^{\tau(\Omega-2)},
\end{equation}

\noindent and the relative change

\begin{equation}\label{DPBlochspeedchange}
   \kappa(t)=-4(2\mathbf{z}+\mathbf{y})e^{-4
(2\mathbf{z}+\mathbf{y})t}
\end{equation}

\noindent of the deformed-Bloch-sphere volume. The von Neumann
entropy and the trace distance are numerically calculated for the
initial state defined by $u=0=v$, eq.\eqref{DensOperBloch}.
Comparison with the standard DP channel follows from placing
$\mathbf{x}=0$ and $\mathbf{y}=\mathbf{z}$ in
eq.\eqref{DPDensOperBlochxyz}.

\begin{figure}
\centering $\begin{array}{cc}
 \includegraphics[width=0.35\textwidth]{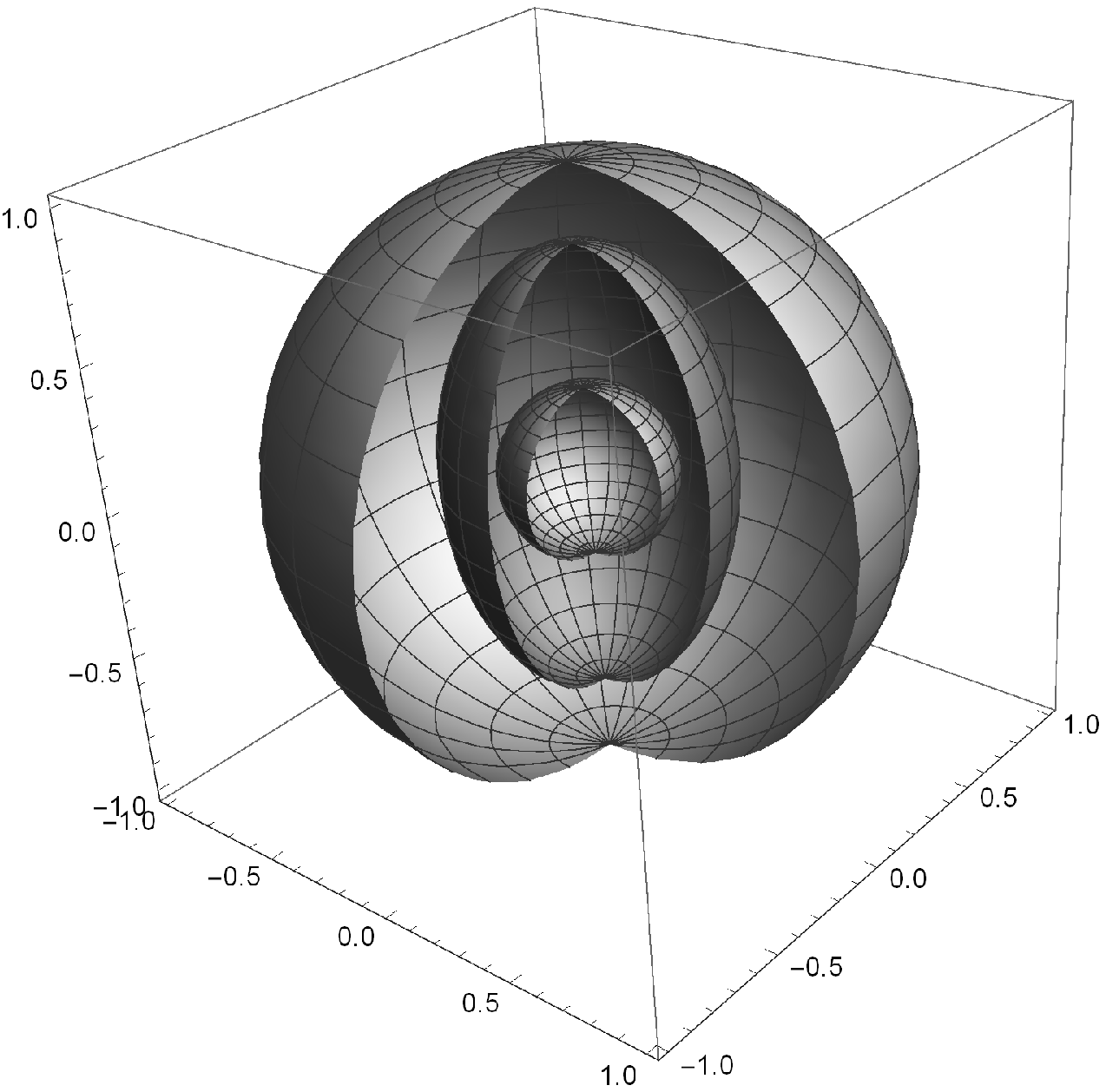}&
  \includegraphics[width=0.4\textwidth]{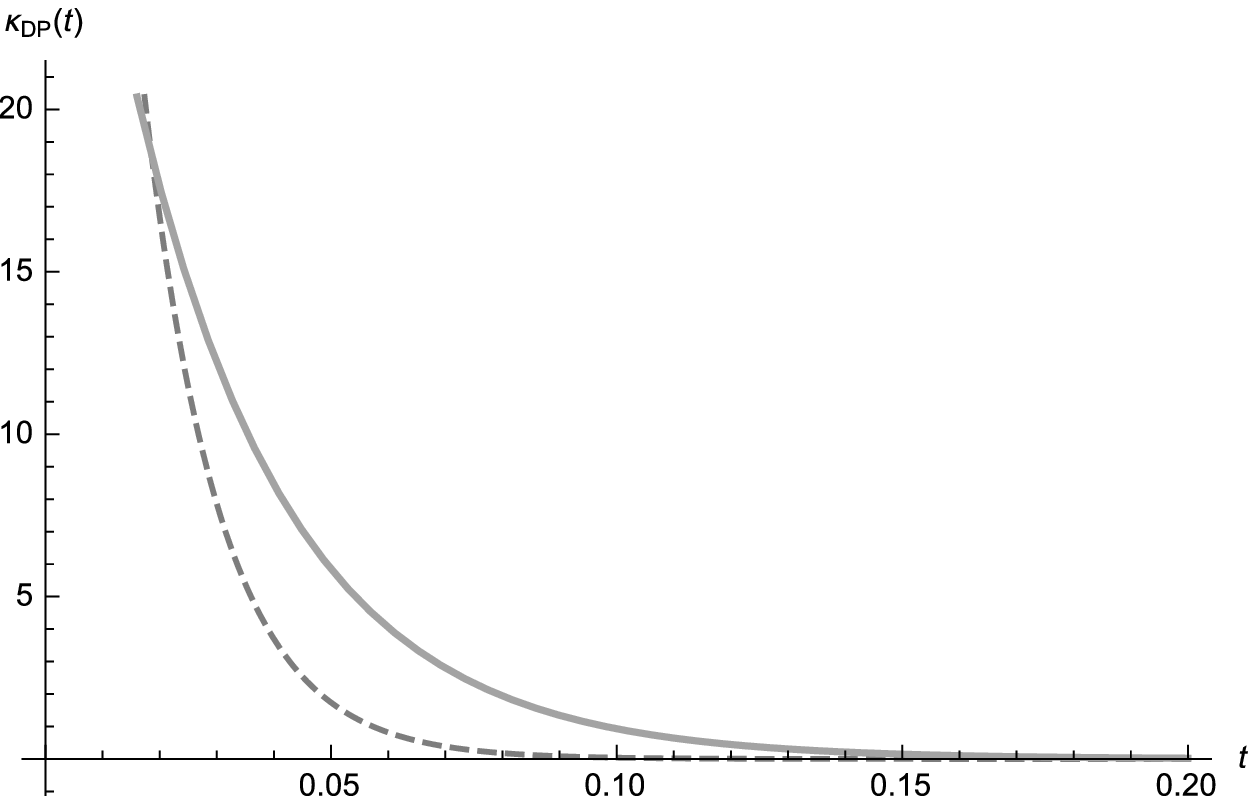}
  \end{array}$
  \caption{(left) The depolarizing channel for: $T=50$,
$\alpha=0.02$, $\omega_0=1$ and $\omega_c=15$. The big sphere is
for $t=0$, the small sphere is obtained for the standard DP
channel ($\mathbf{x}=0$ and $\mathbf{y}=\mathbf{z})$, while the
ellipsoid concerns the generalized DP channel, both for  $t=0.05$.
(Right) The relative change of the Bloch sphere volume. The dashed
line is for the standard DP channel and the thick one for the
generalized DP channel, respectively.} \label{FigVDPRelVDP}
\end{figure}

Fig.\ref{FigVDPRelVDP}(left) exhibits  faster deformation of the
Bloch sphere for the standard than for the generalized DP channel,
while Fig.\ref{FigVDPRelVDP}(right) depicts the volume relative
change, eq.\eqref{DPBlochspeedchange}.

\begin{figure*}
\centering
 $\begin{array}{rl}
    \includegraphics[width=0.4\textwidth]{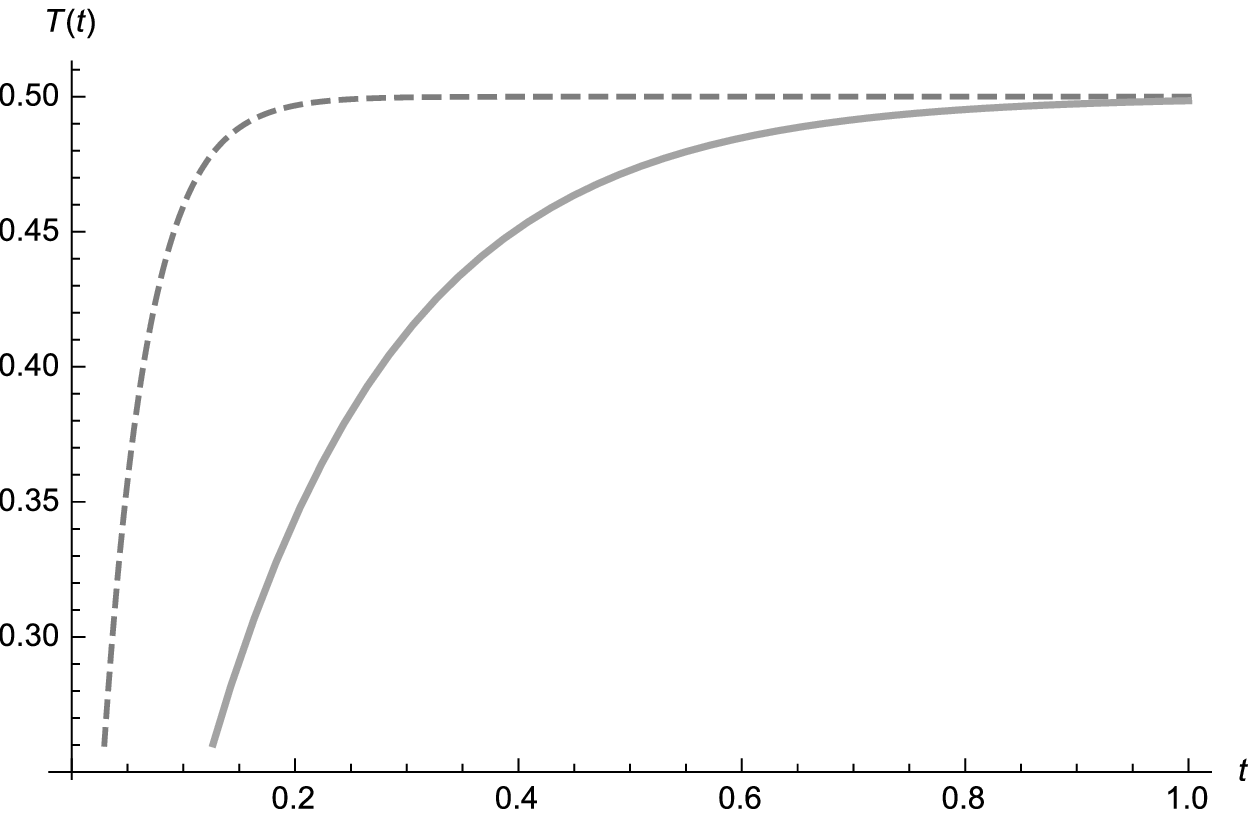} &
    \includegraphics[width=0.4\textwidth]{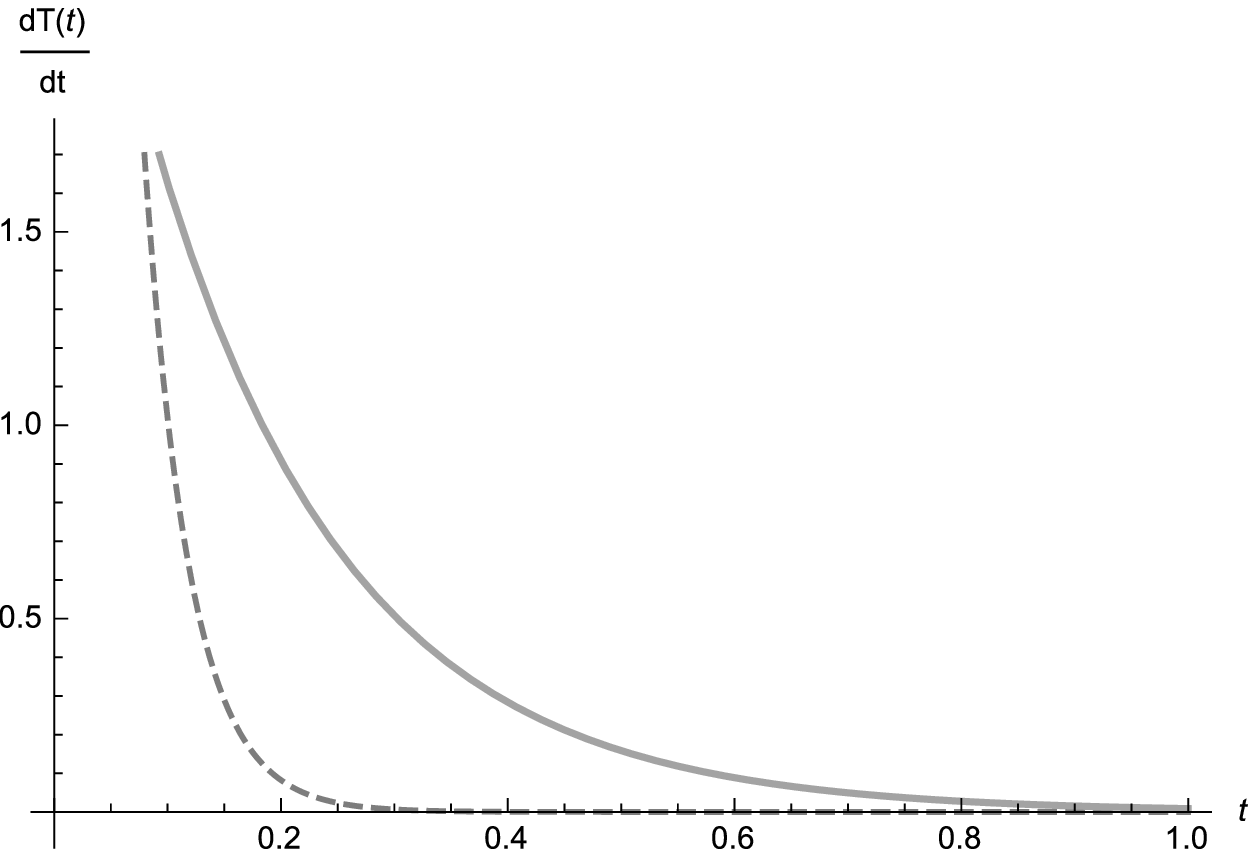}\\
    \multicolumn{2}{c}{\includegraphics[width=0.4\textwidth]{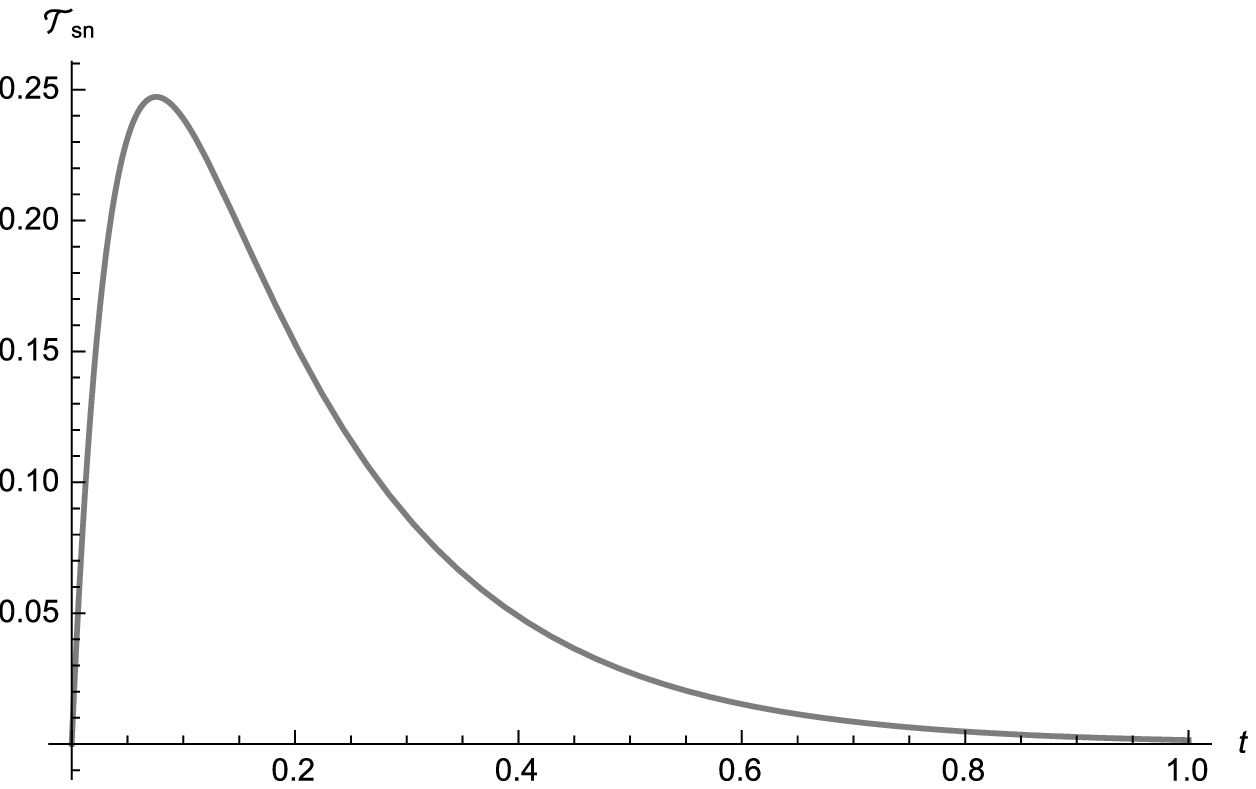}}
\end{array}$
\caption{(Top left) The temporal change of the trace distance;
(Top right) The rate of change of the trace distance; (Bottom) The
trace distance for the states obtained for the two channels. The
dashed lines are for the standard DP channel, while the thick
lines are for the generalized DP channel. The parameters: $T=50$,
$\alpha=0.02$, $\omega_0=1$ and
$\omega_c=15$.}\label{FigChangeTrDistDP}
\end{figure*}

From Fig.\ref{FigChangeTrDistDP}(top left) we can see that, for
long times, the trace distance  tends to be  equal for both
channels.

\begin{figure}
\centering $\begin{array}{cc}
  \includegraphics[width=0.4\textwidth]{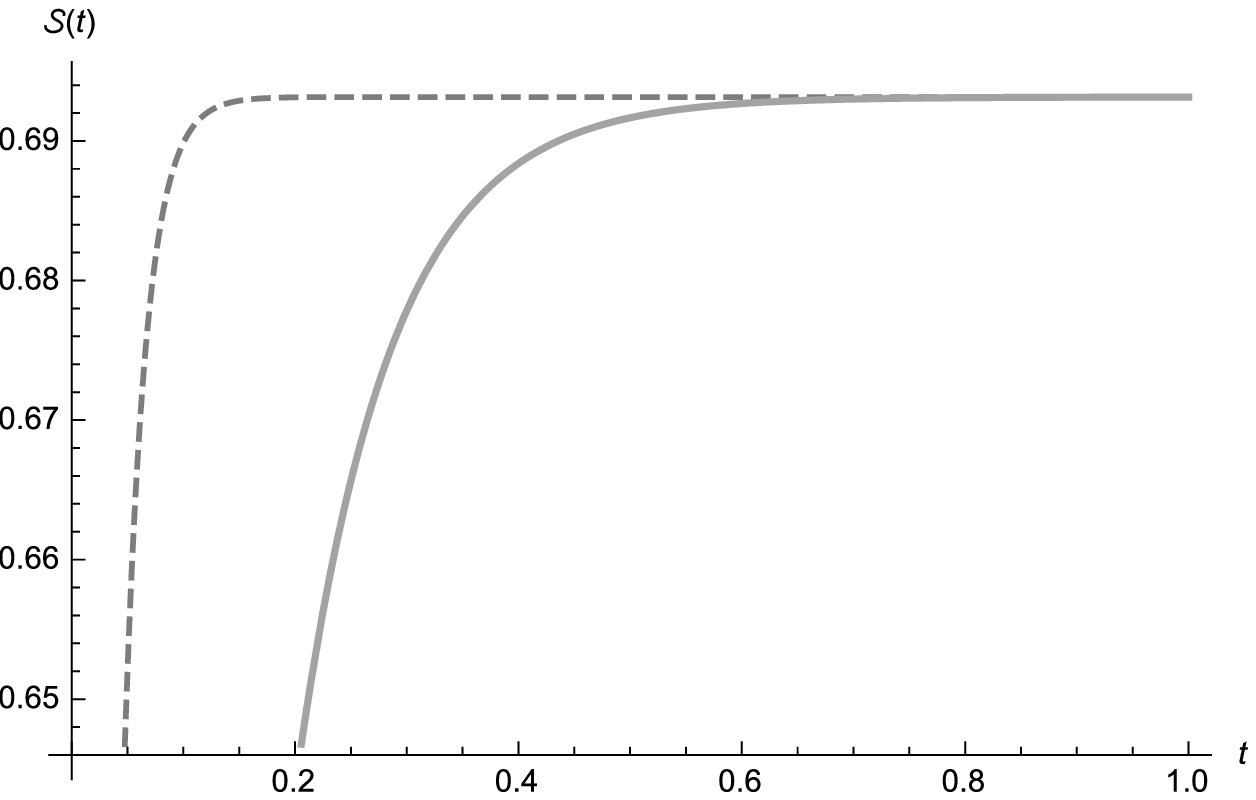}&
   \includegraphics[width=0.4\textwidth]{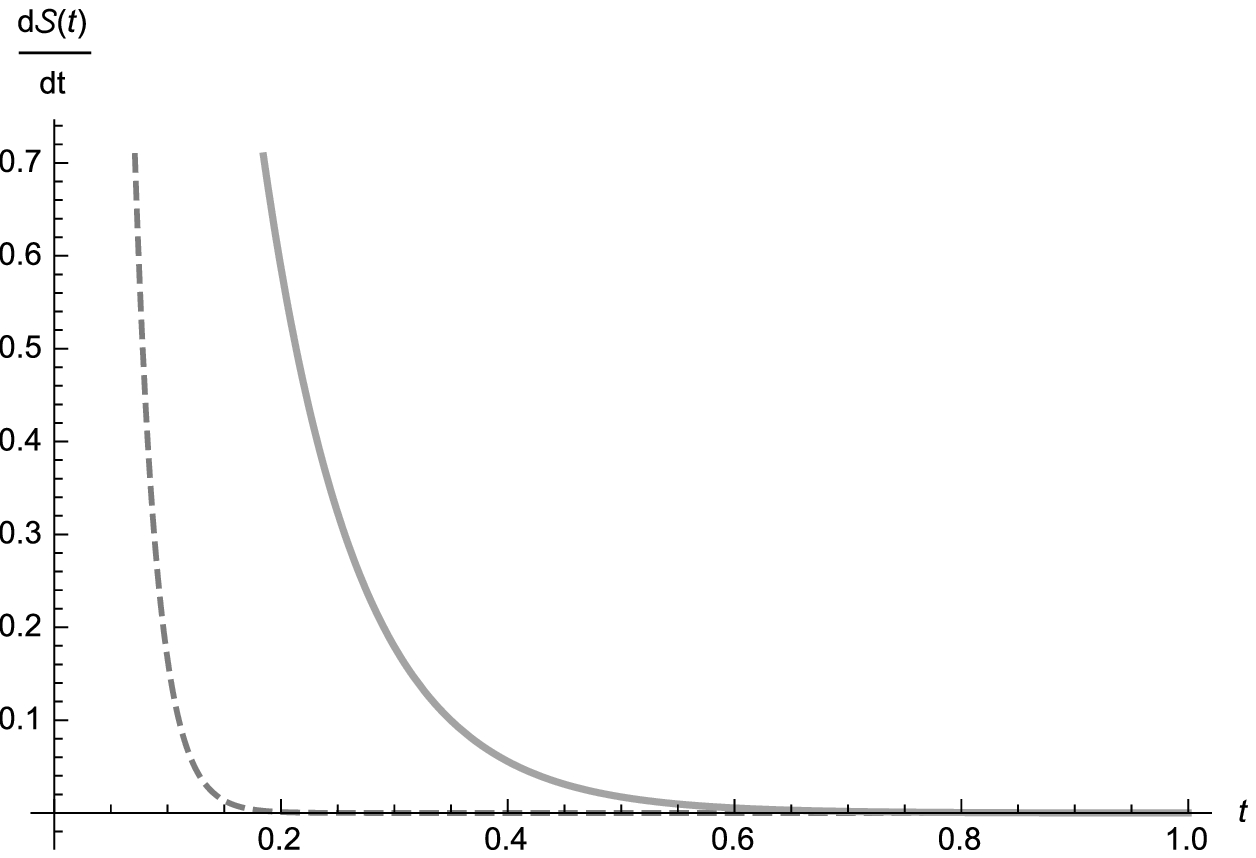}
   \end{array}$
\caption{(Left) The entropy  change (both lines start from zero,
which is not shown here)(Right) The rate of the entropy change.
The dashed lines are for the standard DP channel, while the thick
lines are for the generalized DP channel. The parameters: $T=50$,
$\alpha=0.02$, $\omega_0=1$ and $\omega_c=15$.} \label{FigEntroDP}
\end{figure}

Time dependence of the entropy and the rate of its change are
depicted on Fig.\ref{FigEntroDP}.

\bigskip

\begin{figure}
\centering
  \includegraphics[width=0.35\textwidth]{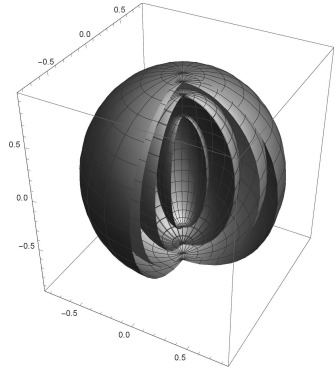}

\caption{The GDP process ellipsoids for different parameters
values, starting from the largest one, respectively: $T=50,
\alpha=0.005, \omega_c=15$; $T=100, \alpha=0.005, \omega_c=50$;
$T=50, \alpha=0.02, \omega_c=15$; $T=50, \alpha=0.02,
\omega_c=50$; $T=100, \alpha=0.02, \omega_c=15$. The parameters
$\omega_0=1$ and $t=0.05$ are kept fixed. The ellipsoids
pertaining to ($T=50, \alpha=0.02, \omega_c=15$) and  ($T=50,
\alpha=0.02, \omega_c=50$) are mutually almost indistinguishable.}
\label{FigDP3DComparison}
\end{figure}

\begin{figure*}
\centering
 $\begin{array}{rl}
    \includegraphics[width=0.4\textwidth]{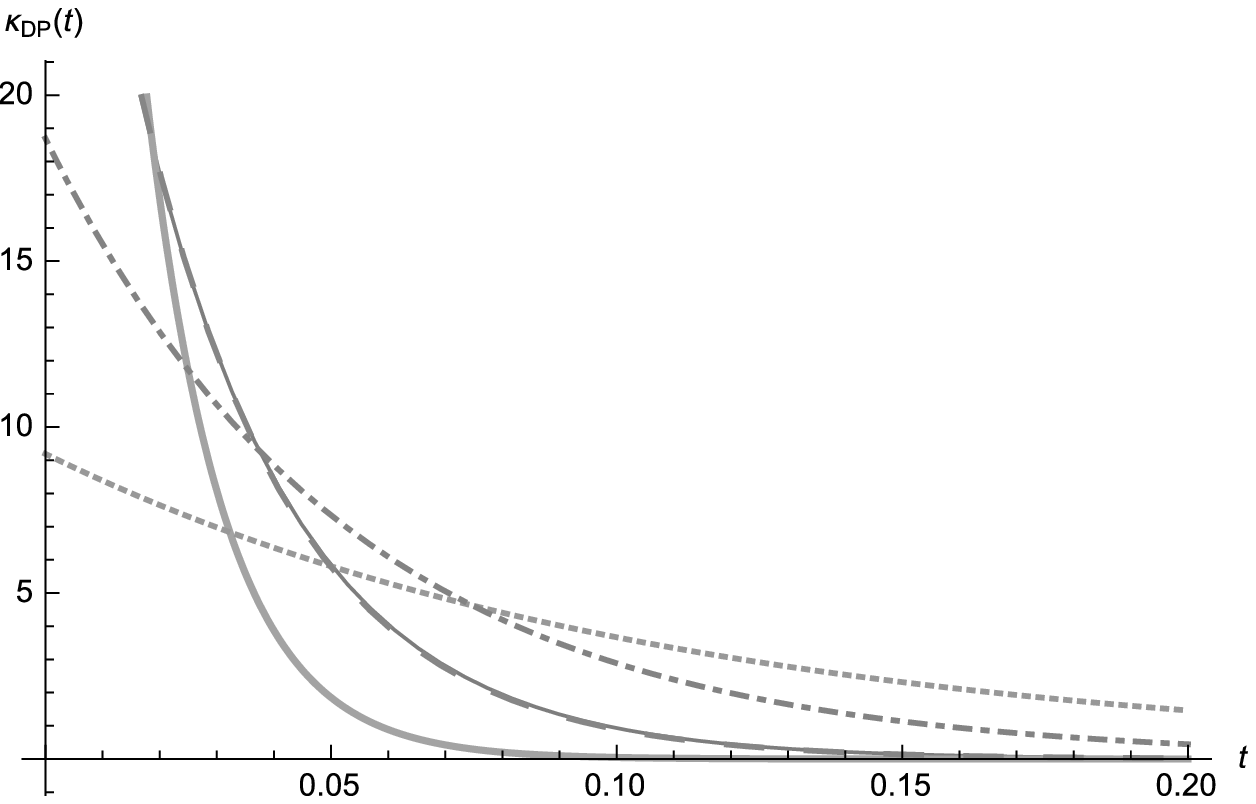} &
    \includegraphics[width=0.4\textwidth]{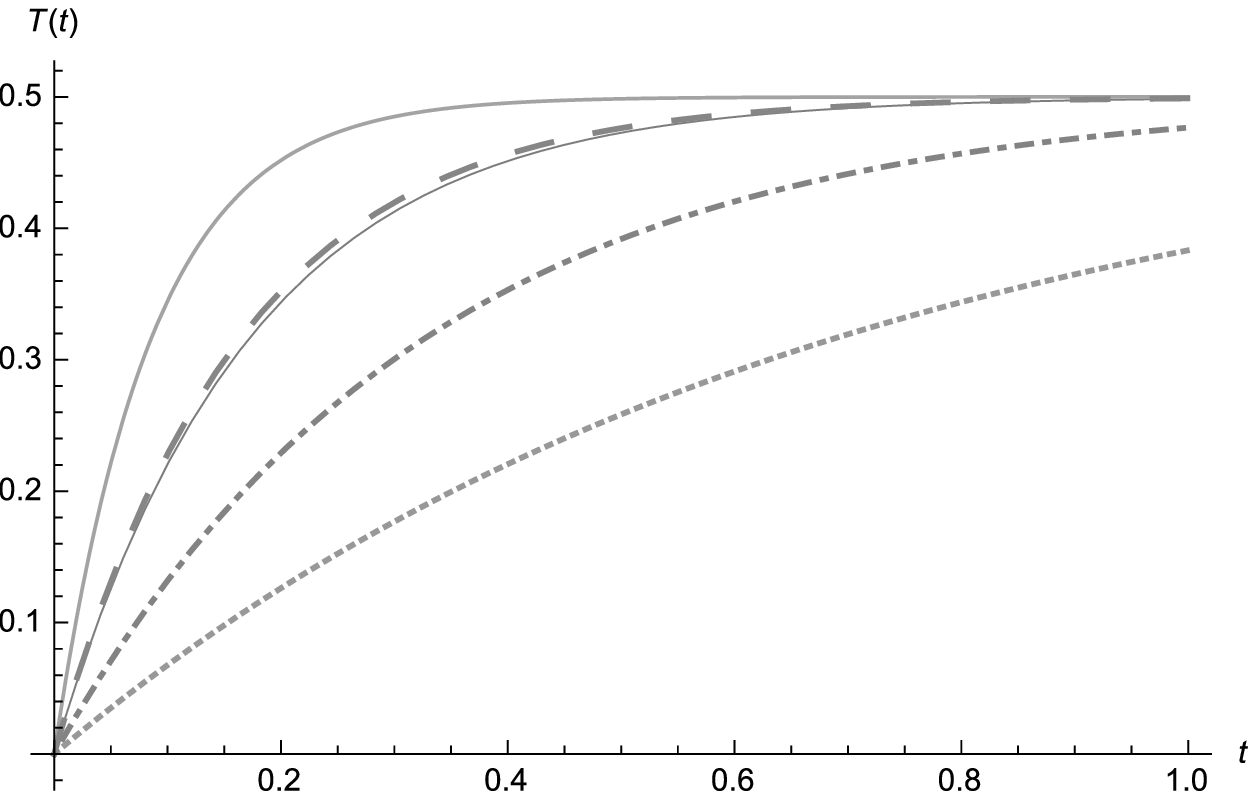}\\
    \multicolumn{2}{c}{\includegraphics[width=0.4\textwidth]{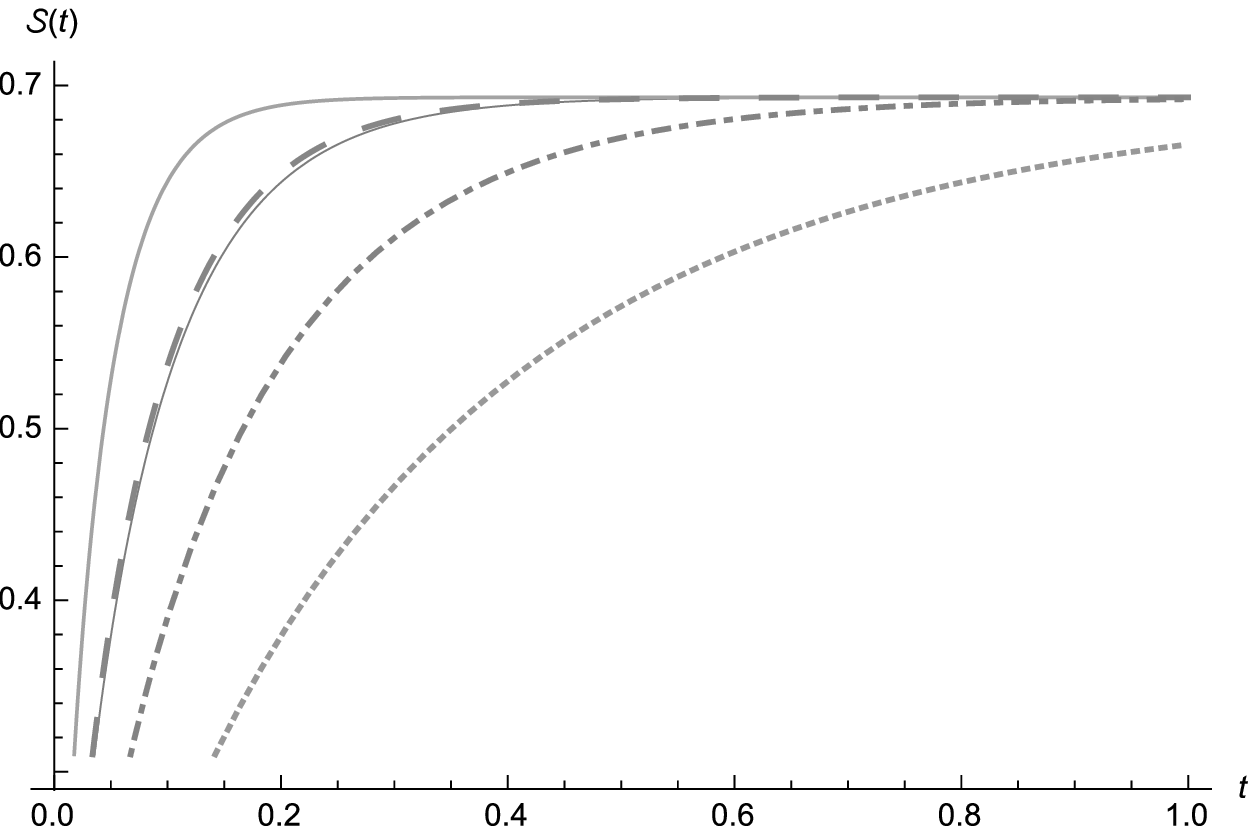}}
\end{array}$
\caption{(Top left) The relative change of the Bloch sphere
volume. (Top right) The trace-distance change. (Bottom) The
entropy change. For all figures the meaning of the lines is the
same: thinner dashed line  ($T=50$, $\alpha=0.005$ ,
$\omega_c=15$), dot-dashed line  ($T=100$, $\alpha=0.005$ i
$\omega_c=50$), thinner  solid  line ($T=50$, $\alpha=0.02$ i
$\omega_c=15$), thicker dashed line ($T=50$, $\alpha=0.02$ i
$\omega_c=50$), thicker solid line ($T=100$, $\alpha=0.02$ i
$\omega_c=15$).  The parameter $\omega_0=1$ is kept fixed.
}\label{FigComparison}
\end{figure*}

\bigskip

Expectably, operation of the GDP channel is sensitive to the
values of the parameters $T$, $\alpha$ and $\omega_c$. Qualitative
similarity of the GDP-channel operations for certain values of the
parameters is depicted in Figs. \ref{FigDP3DComparison} and
\ref{FigComparison}.

From Figs. \ref{FigDP3DComparison} and \ref{FigComparison}
increase in temperature $T$ and the coupling constant $\alpha$
gives rise to the more efficient action of the channel. To this
end, relatively the most relevant contribution is due to the
strength of the interaction-which, of course, cannot be
arbitrarily large for the weak-coupling Markovian processes such
as the GDP channel. The contribution of the cutoff frequency comes
essentially from the better satisfied formal  condition for
Markovianity, i.e. form the inequality $\omega_0/\omega_c \ll 1$,
and hence is of the secondary importance.

\section{Discussion}\label{Discussion}

Microscopic derivation of the Kraus operators is poorly addressed
in the literature. Nevertheless, microscopically derived Kraus
operators provide the physically richer and mathematically the
more accurate description of the realistic physical processes.
Hence we can expect that the parameters control can in principle
give rise to the open system's partial control. Bearing in mind
that the error-correction protocols \cite{N&Ch} are adapted to the
concrete forms of the Kraus operators, the knowledge of the exact,
generalized Kraus operators eqs.\eqref{DP1}-\eqref{DP4} can help
in devising the more reliable and more accurate error-correction
protocols
\cite{Terhal,Raussendorf,ShorPreskill,KrausGisinRen,BBPSSW}. To
this end, the results will be presented elsewhere. Here we just
emphasize that our results regard the short-time intervals of
interest for the realistic quantum information/computation
protocols/algorhithms.

Practical implementation of the quantum information-theoretic
protocols \cite{N&Ch} {\it assumes} that the quantum ``hardware''
operates as it is theoretically described. This non-trivial
assumption is here theoretically addressed in the context of the
one-qubit quantum noise channels. The results regarding the
amplitude damping and the phase damping quantum channels are
presented in \cite{KJS}.

On the other hand, to the best of our knowledge, the generalized
Kraus operators for the depolarizing channel, eqs.
\eqref{DP1}-\eqref{DP4}, are here derived for the first time
exhibiting a subtle yet physically important dissent from the
standard and widely used ones, eq.\eqref{RomLoFrancoDPKraus}. On
this basis we can recognize the following research lines of
interest.

From Figs. \ref{FigDP3DComparison} and \ref{FigComparison} we can
learn that higher temperature and strong interaction give rise to
the more efficient and faster action of the generalized
depolarizing channel that, at least in principle, can be
experimentally tested. On the other hand, transition to the
Schr\"{o}dinger picture:
\begin{equation}\label{SchredKraus}
   \hat{\mathbb{E}}_i^S (t)= \hat{U}(t)\hat{\mathbb{E}}_i (t),
\end{equation}
where $\hat{U}(t)= e^{-\frac{i}{\hbar}\hat{H}_qt}$ and $\hat{H}_q$
is the qubit's self-Hamiltonian, provides a study of the
external-fields influence on the single-qubit's dynamics.

Temporal change of the Bloch sphere, Section \ref{Analysis},
reveals that the standard channel is  more detrimental than the
here introduced generalized channel for the short time intervals.

However, for long times we observe the full match of the standard
and the here derived generalized DP channel. This can be seen in
the asymptotic limit ($\tau\to\infty$) as follows. On the one
hand, the state in eq.\eqref{DPDensOperBlochthetaOmegatau} clearly
satisfies $\lim_{\tau\to\infty} \phi_{\tau}(\hat{\rho}) =
\hat{I}/2$, which is the defining feature of the standard DP
channel, eq.\eqref{RomLoFrancoDPKraus}. On the other hand, in the
asymptotic limit, all the Kraus operators,
eqs.\eqref{DP1}-\eqref{DP4}, \eqref{DP1Stand}-\eqref{DP4Stand} and
eq.\eqref{RomLoFrancoDPKraus}, give the unique set of Kraus
operators:

\begin{equation}\label{DPasym1}
\mathbb{E}_1=\frac{1}{2}\left(
\begin{array}{cc}
 0 & - i  \\
  i  & 0
\end{array}
\right),
\end{equation}

\begin{equation}\label{DPasym2}
\mathbb{E}_2=\frac{1}{2}\left(
\begin{array}{cc}
 0 &1 \\
 1 & 0
\end{array}
\right),
\end{equation}

\begin{equation}\label{DPasym3}
\mathbb{E}_3=\frac{i}{2}\left(
\begin{array}{cc}
1 & 0 \\
 0 & -1
\end{array}
\right),
\end{equation}

\begin{equation}\label{DPasym4}
\mathbb{E}_4= \frac{i}{2}\left(
\begin{array}{cc}
1 & 0 \\
 0 & 1
\end{array}
\right),
\end{equation}

\noindent that obviously satisfy the completeness relation.

With the use of eqs.\eqref{A12}-\eqref{A14}, the condition $2
\gamma_{zz}(0,T)=\gamma_{xx}(\omega_0,T)+\gamma_{yy}(\omega_0,T)$
(Section \ref{KrausDP}) for reducing the GDP to the DP channel,
numerically gives the unique (physically relevant, i.e.
non-negative)  value $\omega_{\circ} = 0$ for every temperature
$T$. Then neglecting the Lamb-shift term makes the
interaction-picture equation \eqref{RomLoFrancoDPmaster} valid
also in the Schr\" odinger picture.

As an application, we consider entanglement dynamics for a pair of
qubits regarding the following typical scenario in quantum
information-processing. Alice and Bob share a pair of the
initially entangled qubits (denoted $1$ and $2$) and locally and
independently perform operations on their respective qubits. The
qubits are sufficiently remote from each other and hence do not
mutually interact while being subjected to the mutually
independent, local GDP channels. The task is to investigate
dynamics of entanglement for the pair $1+2$.

Being spatially remote, the qubits are dynamically independent.
Therefore the Kraus operators for the combined $1+2$ system are
the tensor-product Kraus operators for the individual qubits.
Transition to the Schr\" odinger picture is fulfilled by the
unitary operations, $U_i(t)=\exp(-\imath t H_{i}/\hbar)$ for the
qubits $i=1,2$, independently of each other. Given the
combined-system's state (in the Schr\" odinger picture):
\begin{equation}
\rho(t) = \sum_{i,j} U_1(t)K_{1i}(t)\otimes U_{2}(t) K_{2j}(t)
\vert \Psi(0)\rangle\langle\Psi(0)\vert  K_{1i}^{\dag}
U_1^{\dag}(t)\otimes K_{2j}^{\dag}(t) U_2^{\dag}(t),
\end{equation}

\noindent we  calculate the so-called concurrence for a
bipartite mixed state $\rho(t)$ that is defined as \cite{Wooters}:
\begin{equation}
C(\rho(t)) = \max \{0, \Lambda(t)\},
\end{equation}

\noindent where

\begin{equation}
\Lambda(t) = \sqrt{\lambda_1(t) }- \sqrt{\lambda_2(t) }- \sqrt{\lambda_3(t) }
- \sqrt{\lambda_4(t) },
\end{equation}

\noindent with the eigenvalues $\lambda_1
> \lambda_2 > \lambda_3 > \lambda_4$ of

\begin{equation}\label{VutersRho}
\rho(t) (\sigma_{1y}\otimes\sigma_{2y}) \rho^{\ast}(t)
(\sigma_{1y}\otimes\sigma_{2y}).
\end{equation}

\noindent In \eqref{VutersRho}, $\sigma_{iy}$ is the $y$-Pauli-matrix for the
$i$th qubit, while $\ast$ denotes the standard complex
conjugation. For the initial entangled state we choose
$\vert\Psi\rangle = [\vert 0\rangle_1\vert 0\rangle_2 + \vert 1\rangle_1 \vert 1\rangle_2
]/\sqrt{2}$, where we omit the symbol of the tensor-product and $\vert i\rangle, i=0,1$ represent the eigenbasis of the $\sigma_z$ Pauli matrix.

\begin{figure}
\centering
 \includegraphics[width=0.4\textwidth]{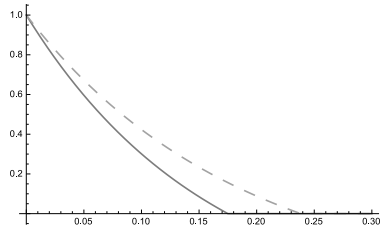}
  \caption{Concurrence dynamics for the standard DP channel (solid line) and the GDP channel (dotted line). The parameters: $\omega_{1}=0.1$, $\omega_{2}=0.2$, $T=10$, $\alpha=0.02$.} \label{CcDPandGDP}
\end{figure}

Bearing in mind the definition of the so-called entanglement-of-formation \cite{Wooters}:
\begin{equation}
\mathcal{E}(\rho) = H\left({1+\sqrt{1 + C^2(t)}\over 2}\right)
\end{equation}

\noindent where $H(x) = -x \log_2 x - (1-x) \log_2(1-x)$, from Fig. \ref{CcDPandGDP}, we conclude about the occurrence of the ``entanglement sudden death" \cite{Yu Eberly} also for the GDP channel. That is, for the concurrence $C(t)=0$ for some $t$, the entanglement of formation $\mathcal{E}=0$ in the same instant of time--the loss of the initial entanglement in the instant of time $t$. Fig. \ref{CcDPandGDP} justifies the above finding that the GDP channel is less deteriorating (slower occurrence of the entanglement sudden death) than the standard DP channel.

\section{Conclusion}\label{Conclusion}

Microscopic derivation of the Kraus operators is a new emerging
line of research of foundational as well as of interest for
application in modern open quantum systems theory. The
microscopically derived Kraus operators for the generalized
depolarizing channel are here presented for the first time while
carrying the clear signatures of the microscopic-models details.
Implications for the possible open-system's control and for an
elaboration of the known error-correction protocols will be
presented elsewhere.

\appendix
\section{Derivation of the generalized depolarizing
master equation}\label{Appendix}

\bigskip

\noindent Here we derive a  Markovian master equation, which
describes a completely positive and trace preserving, homogeneous
Markovian process for the microscopic model eq.\eqref{A1} in the
body text. Such processes are known to be presented by a master
equation of the general Lindblad form \cite{RivasHuelga,BreuPetr}:

\begin{equation}\label{A2}
   \dot{\hat{\rho}}_{S}(t)=-i[\hat{H}_{LS},\hat{\rho}_S(t)]+ \sum_{\omega}
\sum_{k,l}\gamma_{kl}(\omega)[\hat{A}_{l}(\omega)\hat{\rho}_{S}(t)\hat{A}_{k}^{\dag}
(\omega)-\frac{1}{2}\{\hat{A}_{k}^{\dag}(\omega)\hat{A}_{l}(\omega),\hat{\rho}_{S}(t)\}],
\end{equation}

\noindent where the Lamb shift:

\begin{equation}\label{A3}
\hat{H}_{\mathrm{LS}}=\sum_\omega\sum_{k,\ell}S_{k\ell}
(\omega)\hat{A}_k^\dagger(\omega)\hat{A}_\ell(\omega),
\end{equation}

\noindent and the correlation (damping) functions
$\gamma_{kl}(\omega)$:

\begin{equation}\label{A4}
   \gamma_{kl}(\omega)=2\pi\mathrm{tr}\left[\hat{B}_k(\omega)
   \hat{B}_\ell\hat{\rho}_\mathrm{Eth}\right],
\end{equation}

\noindent while $\hat{\rho}_{Eth}$ is the environment's thermal
equilibrium state, $\hat\rho_{Eth} = \exp(-\beta \hat H_E)/Z$, on
the inverse temperature $\beta = 1/k_B T$.

With the definition \cite{RivasHuelga}:

\begin{equation}\label{A5}
   \hat{A}_{k}(\omega)=\sum_{\varepsilon'-\varepsilon=\omega}|\psi_\varepsilon\rangle\langle\psi_
\varepsilon|\hat{A}_{k}|\psi_{\varepsilon'}\rangle\langle\psi_{\varepsilon'}|,
\end{equation}

\noindent where $|\psi _\varepsilon\rangle$ are eigenstates of the
system's self-Hamiltonian, $\hat H_S$, and $\hat{A}_k$ appearing
in eq.\eqref{A1} of the body text, and analogously for
$\hat{B}_{k}(\omega)$, $k=x,y,z$,  we obtain:

\begin{subequations}\label{A6}
\begin{align}
\hat{A}_1(0)&=\hat{\sigma}_z,  &  \hat{A}_1(\omega_0)&=0,
&\hat{A}_1(-\omega_0)&=0,\\
\hat{A}_2(0)&=0, & \hat{A}_2(\omega_0)&=\hat{\sigma}_{-},
&\hat{A}_2(-\omega_0)&=\hat{\sigma}_{+},\\
\hat{A}_3(0)&=0, & \hat{A}_3(\omega_0)&=i \hat{\sigma}_{-},
&\hat{A}_3(-\omega_0)&=-i \hat{\sigma}_{+} \,.
\end{align}
\end{subequations}

From  the second term in eq. \eqref{A2} and having in mind
eq.\eqref{A6}, after some algebra we obtain the dissipator,
$\mathcal{D}(\hat{\rho}_{S})$:

\begin{equation} \label{A7}
\begin{split}
\mathcal{D}(\hat{\rho}_{S})&=
\gamma_{zz}(0)[\hat{\sigma}_z\hat{\rho}_S\hat{\sigma}_z-\hat{\rho}_S]+
\gamma_{yy}(\omega_0)[\hat{\sigma}_{-}\hat{\rho}_{S}\hat{\sigma}_{+}-
\displaystyle\frac{1}{2}\{\hat{\sigma}_{+}\hat{\sigma}_{-},\hat{\rho}_S\}]+ \\
  & + \gamma_{yx}(\omega_0)[i \hat{\sigma}_{-}\hat{\rho}_{S}\hat{\sigma}_{+}-
\frac{i}{2}\{\hat{\sigma}_{+}\hat{\sigma}_{-},\hat{\rho}_S\}]+\gamma_{xy}(\omega_0)
[-i \hat{\sigma}_{-}\hat{\rho}_{S}\hat{\sigma}_{+}+
\frac{i}{2}\{\hat{\sigma}_{+}\hat{\sigma}_{-},\hat{\rho}_S\}]+ \\
  & + \gamma_{xx}(\omega_0)[\hat{\sigma}_{-}\hat{\rho}_{S}\hat{\sigma}_{+}-
                             \frac{1}{2}\{\hat{\sigma}_{+}\hat{\sigma}_{-},\hat{\rho}_S\}]+
\gamma_{yy}(-\omega_0)[\hat{\sigma}_{+}\hat{\rho}_{S}\hat{\sigma}_{-}-
\frac{1}{2}\{\hat{\sigma}_{-}\hat{\sigma}_{+},\hat{\rho}_S\}]+ \\
  & + \gamma_{yx}(-\omega_0)[-i
\hat{\sigma}_{+}\hat{\rho}_{S}\hat{\sigma}_{-}+
\frac{i}{2}\{\hat{\sigma}_{-}\hat{\sigma}_{+},\hat{\rho}_S\}]+ \\
& + \gamma_{xy}(-\omega_0)[i
\hat{\sigma}_{+}\hat{\rho}_{S}\hat{\sigma}_{-}-
\frac{i}{2}\{\hat{\sigma}_{-}\hat{\sigma}_{+},\hat{\rho}_S\}] +
\gamma_{zz}(-\omega_0)[\hat{\sigma}_{+}\hat{\rho}_{S}\hat{\sigma}_{-}-
\frac{1}{2}\{\hat{\sigma}_{-}\hat{\sigma}_{+},\hat{\rho}_S\}] \,,
\end{split}
\end{equation}

where
$\hat{\sigma}_{\pm}=\frac{1}{2}(\hat{\sigma}_x\pm\imath\hat{\sigma}_y)$.

In accordance with eq.\eqref{A1}, the initial thermal state of the
 bath $E$ can be chosen:
\begin{equation}\label{A8}
   \hat{\rho}_{Eth}(0)=\hat{\rho}_{Ex}\otimes\hat{\rho}_{Ey}\otimes\hat{\rho}_{Ez},
\end{equation}
where the thermal states $\hat{\rho}_{Ei}=\exp(-\beta
\hat{H}_{Ei}){/}Z_{Ei}$, $i=x,y,z$, and $\sum_i \hat{H}_{Ei} =
\hat{H}_E$, for the inverse temperature $\beta$. That is, the
qubit's environment $E$ can be regarded as formally consisting of
three noninteracting subsystems (modes) so that these modes,
represented as the independent environments $E_i$, are
independently coupled to the system's $\hat{\sigma}_x$,
$\hat{\sigma}_y$ and $\hat{\sigma}_z$ Pauli operators. Then it
easily follows that the cross terms in eq.\eqref{A7},
$\gamma_{ij}$, $i\neq j=x,y,z$, fall-off as a direct consequence
of eq.\eqref{A4}. Now, for noninteracting $E_i$ environments, the
dissipator  eq.\eqref{A7} reads:

\begin{equation} \label{A9}
\begin{split}
\mathcal{D}(\hat{\rho}_{S})&=
\gamma_{zz}(0)[\hat{\sigma}_z\hat{\rho}_S\hat{\sigma}_z-\hat{\rho}_S]+
\gamma_{yy}(\omega_0)[\hat{\sigma}_{-}\hat{\rho}_{S}\hat{\sigma}_{+}-
\frac{1}{2}\{\hat{\sigma}_{+}\hat{\sigma}_{-},\hat{\rho}_S\}]+ \\
  & + \gamma_{xx}(\omega_0)[\hat{\sigma}_{-}\hat{\rho}_{S}\hat{\sigma}_{+}-
                             \frac{1}{2}\{\hat{\sigma}_{+}\hat{\sigma}_{-},\hat{\rho}_S\}]+
\gamma_{yy}(-\omega_0)[\hat{\sigma}_{+}\hat{\rho}_{S}\hat{\sigma}_{-}-
\frac{1}{2}\{\hat{\sigma}_{-}\hat{\sigma}_{+},\hat{\rho}_S\}]+ \\
  & + \gamma_{xx}(-\omega_0)[\hat{\sigma}_{+}\hat{\rho}_{S}\hat{\sigma}_{-}-
\frac{1}{2}\{\hat{\sigma}_{-}\hat{\sigma}_{+},\hat{\rho}_S\}]\,.
\end{split}
\end{equation}

The model eq.\eqref{A1} and eq.\eqref{A8} limit the choice of the
parameters in order to obtain Markovian dynamics eq.\eqref{A2} as
follows \cite{RivasHuelga}. First, it is required that the
so-called spectral density, denoted $J(\omega)$, be of the Ohmic
kind. Second, the inequality $\omega_0{/}\omega_c \ll 1$ is
required, where $\omega_c$ is the cutoff frequency. Finally,  the
high temperature limit is required for the depolarizing process,
formally defined by $k_BT{/}\hbar\omega_0 \gg 1$.

From eq.\eqref{A4}, the damping function $\gamma_{zz}(0)$:
\begin{equation}\label{A10}
    \gamma_{zz}(0)=2\pi tr[\hat{B}_z(0)\hat{B}_z
    \hat{\rho}_{Eth}(0)],
\end{equation}

\noindent where

\[
\hat{B}_z=\int_{-\omega_\textrm{max}}^{\omega_\textrm{max}}d\omega
\hat{B}_z(\omega), \text{ with}\left\{
\begin{array}{l}
\hat{B}_z(\omega)=h(\omega)a_\omega, \\
\hat{B}_z(-\omega)=h(\omega) a_\omega^\dagger,
\end{array}
\right. \text{for }\omega>0.
\]

Since $\hat{B}_z(\omega=0)$ is not well-defined, we stick to
\cite{RivasHuelga}:
\begin{equation}\label{A11}
    \gamma_{zz}(0)=2\pi \lim_{\omega\rightarrow0} tr[\hat{B}_z(0)\hat{B}_z
    \hat{\rho}_{Eth}(0)],
\end{equation}

\noindent with the often used Ohmic spectral density
$J(\omega)=\alpha\omega e^{-\frac{\omega}{\omega_c}}$ thus
obtaining:

\begin{equation}\label{A12}
    \gamma_{zz}(0)=2\pi\lim_{\omega\rightarrow0}
J(|\omega|)\langle n(|\omega|)\rangle,
\end{equation}
while
\begin{equation}\label{A13}
    \gamma_{yy}(\omega_0)=\frac{\pi}{2}J(\omega_0)[\langle n(\omega_0)\rangle+1]
\end{equation}
and
\begin{equation}\label{A14}
     \gamma_{yy}(-\omega_0)=\frac{\pi}{2}J(\omega_0)\langle
     n(\omega_0)\rangle,
\end{equation}
and analogously for $\gamma_{xx}$s; the constant $\alpha$ is the
interaction strength, i.e. the weak-coupling constant.

It should be stressed  here  that, for high temperature, $\langle
n(\omega_0)\rangle=\left[e^{(\omega_0/T)}-1\right]^{-1}\gg 1$ and
thus:
\begin{equation}\label{A15}
    \gamma_{yy}(\omega_0)\approx \gamma_{yy}(-\omega_0)=\frac{\pi}{2}
    J(\omega_0)\langle n(\omega_0)\rangle,
\end{equation}
and analogously for $\gamma_{xx}$s. Hence the dissipator
eq.\eqref{A9} simplifies:
\begin{equation} \label{A16}
\begin{split}
\mathcal{D}(\hat{\rho}_{S})&=
\gamma_{zz}(0)[\hat{\sigma}_z\hat{\rho}_S\hat{\sigma}_z
-\hat{\rho}_S]+\\&
+\gamma_{yy}(\omega_0)[\hat{\sigma}_{-}\hat{\rho}_{S}\hat{\sigma}_{+}+
\hat{\sigma}_{+}\hat{\rho}_{S}\hat{\sigma}_{-}-
\frac{1}{2}\{\hat{\sigma}_{+}\hat{\sigma}_{-},\hat{\rho}_S\}-
 \\& -\frac{1}{2}\{\hat{\sigma}_{-}\hat{\sigma}_{+},\hat{\rho}_S\}]
   + \gamma_{xx}(\omega_0)[\hat{\sigma}_{-}\hat{\rho}_{S}\hat{\sigma}_{+}+\hat
                             {\sigma}_{+}\hat{\rho}_{S}\hat{\sigma}_{-}-\\&
                             -\frac{1}{2}\{\hat{\sigma}_{+}\hat{\sigma}_{-},\hat{\rho}_S\}-
\frac{1}{2}\{\hat{\sigma}_{-}\hat{\sigma}_{+},\hat{\rho}_S\}]\,.
\end{split}
\end{equation}

With the  use of the identities:
$\hat{\sigma}_{-}\hat{\rho}\hat{\sigma}_{+}+\hat{\sigma}_{+}\hat{\rho}\hat{\sigma}_{-}=
\frac{1}{2}(\hat{\sigma}_{x}\hat{\rho}\hat{\sigma}_{x}+\hat{\sigma}_{y}\hat{\rho}\hat{\sigma}_{y})$
and
$\{\hat{\sigma}_{+}\hat{\sigma}_{-}\hat{\rho}+\hat{\sigma}_{-}\hat{\sigma}_{+},\hat{\rho}\}=2\hat{\rho}$,
eq.\eqref{A16} reads:
\begin{equation} \label{A17}
\begin{split}
\mathcal{D}(\hat{\rho}_{S})&=\gamma_{zz}(0)
(\hat{\sigma}_{z}\hat{\rho}_S\hat{\sigma}_{z}-\hat{\rho}_S)+\\&
+\frac{\gamma_{yy}(\omega_0)+\gamma_{xx}(\omega_0)}{2}
(\hat{\sigma}_{x}\hat{\rho}_S\hat{\sigma}_{x}-\hat{\rho}_S)+\\&
+\frac{\gamma_{yy}(\omega_0)+\gamma_{xx}(\omega_0)}{2}
+(\hat{\sigma}_{y}\hat{\rho}_S\hat{\sigma}_{y}-\hat{\rho}_S)\,.
\end{split}
\end{equation}

The Lamb shift defined by  eq.\eqref{A3} takes the form

\begin{equation}\label{A18}
\begin{split}
   \hat{H}_{LS}&=S_{11}(0)\hat{I}+S_{22}(\omega_0)\hat{\sigma}_{+}\hat{\sigma}_{-}
   +S_{33}(\omega_0)\hat{\sigma}_{+}\hat{\sigma}_{-}+\\&+
   S_{22}(-\omega_0)\hat{\sigma}_{-}\hat{\sigma}_{+}
   +S_{33}(-\omega_0)\hat{\sigma}_{-}\hat{\sigma}_{+}.
   \end{split}
\end{equation}

Calculation of  $S_{ii}(\pm\omega_0)$ follows via the so called
Sochozki's formulae \cite{RivasHuelga}:

\begin{equation}\label{A19}
   S_{22}(\omega_0)=\frac{1}{4}\mathrm{P.V.}\int_{0}^{\omega_\textrm{max}} d\omega'
J(\omega')\left[\frac{\langle
n(\omega')\rangle+1}{(\omega_0-\omega')}+\frac{\langle
n(\omega')\rangle}{(\omega_0+\omega')}\right],
\end{equation}

and

\begin{equation}\label{A20}
    S_{22}(-\omega_0)=\frac{-1}{4}\mathrm{P.V.}\int_{0}^{\omega_\textrm{max}} d\omega' J(\omega')
    \left[\frac{\langle n(\omega')\rangle+1}{(\omega_0+\omega')}+\frac{\langle
    n(\omega')\rangle}{(\omega_0-\omega')}\right],
\end{equation}

and similarly for $S_{33}(\pm\omega_0)$.

At this point, the following conditions facilitate the analysis:
the spectral density $J(\omega)$  is assumed to be the same for
all interaction-terms in eq.\eqref{A1}, while, for high
temperature,
 $S_{22}(-\omega_0)$
$=-S_{22}(\omega_0)$ and $S_{33}(-\omega_0)$ $=-S_{33}(\omega_0)$.
Hence:

\begin{equation}\label{A21}
   \hat{H}_{LS}=S_{11}(0)\hat{I}+(S_{22}(\omega_0)+
   S_{33}(\omega_0))\hat{\sigma}_z.
\end{equation}

Equations \eqref{A19} and \eqref{A20}, in the high temperature
limit, take the form:
\begin{equation}\label{A22}
   S_{22}(\omega_0)=\frac{\omega_0}{2}\mathrm{P.V.}\int_{0}^{\omega_\textrm{max}} d\omega'
J(\omega')\frac{\langle n(\omega')\rangle}{\omega_0^2-\omega'^2}
\end{equation}
and
\begin{equation}\label{A23}
\begin{split}
    S_{22}(-\omega_0)&=-\frac{\omega_0}{2}\mathrm{P.V.}\int_{0}^{\omega_\textrm{max}} d\omega'
J(\omega')\frac{\langle
n(\omega')\rangle}{\omega_0^2-\omega'^2}=\\&=-S_{22}(\omega_0)\,,
\end{split}
\end{equation}
and analogously for $S_{33}(\pm\omega_0)$.

From eqs.\eqref{A21}-\eqref{A23}, the  Lamb shift finally obtains
the form:

\begin{equation}\label{A24}
\begin{split}
   \hat{H}_{LS}&=S_{11}(0)\hat{I}+(S_{22}(\omega_0)+
   S_{33}(\omega_0))\hat{\sigma}_z=S_{11}(0)\hat{I}+
   \\&
   +\frac{\omega_0}{2}\left(\mathrm{P.V.}\int_{0}^{\omega_\textrm{max}} d\omega'J(\omega')\frac{\langle n(\omega')\rangle}{\omega_0^2-\omega'^2}+
   \mathrm{P.V.}\int_{0}^{\omega_\textrm{max}} d\omega'' J(\omega'')\frac{\langle n(\omega'')\rangle}{\omega_0^2-\omega''^2}\right) \hat{\sigma}_z \equiv S_{11}(0)\hat{I}+\Delta\hat{\sigma}_z\,.
\end{split}
\end{equation}

Collecting all the above results follows the   Markovian master
equation:
\begin{equation}\label{A25}
       \begin{split}
\displaystyle\frac{d\hat{\rho}_S(t)}{dt}&=
-i\Delta[\hat{\sigma}_z,\hat{\rho}_S(t)]+
\gamma_{zz}(0)(\hat{\sigma}_{z}\hat{\rho}_S(t)\hat{\sigma}_{z}-\hat{\rho}_S(t))+\\&
+\frac{\gamma_{xx}(\omega_0)+\gamma_{yy}(\omega_0)}{2}(\hat{\sigma}_{x}\hat{\rho}_S(t)
\hat{\sigma}_{x}-\hat{\rho}_S(t))+\\&
+\frac{\gamma_{xx}(\omega_0)+\gamma_{yy}(\omega_0)}{2}
(\hat{\sigma}_{y}\hat{\rho}_S(t)\hat{\sigma}_{y}-\hat{\rho}_S(t))\,.
\end{split}
\end{equation}
The damping rates  $\gamma_{yy}$ and $\gamma_{xx}$ are of the same
form. Eq.\eqref{A25} is the master equation
eq.\eqref{GeneralizedMJ} in the body text.

\section*{Acknowledgments}
\footnotesize
 The work on this paper is financially supported by
Ministry of Science Serbia under contract no 171028 and in part
for MD by the ICTPSEENET-MTP grant PRJ-09 ``Strings and
Cosmology'' in frame of the SEENET-MTP Network.

\end{document}